\documentclass[prl,twocolumn,showpacs,amsmath,amssymb]{revtex4}
\usepackage{graphicx}

\def\simle{
    \mathrel{\rlap{\raise 0.511ex
        \hbox{$<$}}{\lower 0.511ex \hbox{$\sim$}}}}

\def\simge{
    \mathrel{\rlap{\raise 0.511ex
            \hbox{$>$}}{\lower 0.511ex \hbox{$\sim$}}}}

\begin{document}

\title{Lattice study of exotic $S=+1$ baryon}
\author{Shoichi Sasaki}
\email{ssasaki@phys.s.u-tokyo.ac.jp} 
\affiliation{Department of Physics, University of Tokyo,
  Tokyo 113-0033, Japan}

\begin{abstract}
We propose $S$=+1 baryon interpolating operators, which are based on
an exotic description of the antidecuplet baryon like diquark-diquark-antiquark.
By using one of the new operators,
the mass spectrum of the spin-1/2 pentaquark states is calculated
in quenched lattice QCD at $\beta=6/g^2=6.2$ on a $32^3\times48$
lattice. It is found that the $J^P$ assignment of the lowest $\Theta(uudd \bar s)$ state 
is most likely $(1/2)^-$. 
We also calculate the mass of the charm analog of the $\Theta$ and
find that the $\Theta_c(uudd \bar c)$ state lies much higher than the $DN$ threshold,
in contrast to several model predictions.
\end{abstract}
\maketitle
Recently, LEPS collaboration at Spring-8 has observed a very narrow resonance 
$\Theta^{+}(1540)$ in the $K^{-}$ missing-mass spectrum of the 
$\gamma n \rightarrow nK^{+}K^{-}$ reaction on $^{12}C$~\cite{Nakano:bh}. 
A remarkable observation is its strangeness is $S$=+1,
which means that the observed resonance must contain a strange antiquark. 
Thus, the $\Theta^{+}(1540)$ should be an exotic baryon 
state with the minimal quark content $uudd \bar s$.
This discovery is subsequently confirmed in different reactions
by several other collaborations~\cite{Exp}.
It should be noted, however, that the experimental evidence
for the $\Theta^{+}(1540)$ is not very solid yet since there
are a similar number of negative results to be reported~\cite{Hicks:2004vd}.

Theoretically the existence of such a state was predicted long time ago
by the Skyrme model~\cite{Skyrmion}. 
However, the prediction closest in mass and width with the
experiments was made by Diakonov, Petrov and Polyakov 
using a chiral-soliton model~\cite{Diakonov:1997mm}. 
They predicted that it should be a narrow resonance and stressed that it can be 
detected by experiment because of its narrow width.
In a general group theoretical argument with flavor $SU(3)$, 
$S$=+1 pentaquark state should be a member of 
antidecuplet or higher dimensional representation such as 27-plet or 35-plet.
Both the Skyrme model and the chiral-soliton model predict that the lowest $S$=+1 
state appears in the antidecuplet, $I$=0, and its spin and parity should be $(1/2)^{+}$\cite{{Skyrmion},{Diakonov:1997mm}}.
Experimentally, spin, parity and isospin of the $\Theta^{+}(1540)$ 
are not determined yet. After the discovery of the $\Theta^{+}(1540)$, many model 
studies for the pentaquark state are made with different spin, parity and isospin.

Lattice QCD in principle can 
determine these quantum numbers of the $\Theta^{+}(1540)$, 
independent of such arbitrary model assumptions or the experiments.
We stress that there is substantial progress in lattice study of excited 
baryons recently~\cite{Sasaki:2003xc}.
Especially, the negative parity nucleon $N^*(1535)$, which lies close
to the $\Theta^{+}(1540)$, has become an established state 
in quenched lattice QCD~\cite{{Sasaki:2003xc},{Sasaki:2001nf}}.
Here we report that quenched lattice QCD is capable of studying the $\Theta^{+}(1540)$ as well.

Indeed, it is not so easy to deal with the $qqqq{\bar q}$ state 
rather than usual baryons ($qqq$) and mesons ($q \bar q$) in lattice QCD.
The $qqqq{\bar q}$ state can be decomposed into a pair of color singlet states
as $qqq$ and $q \bar q$, in other words, it can decay into 
two-hadron states
even in the quenched approximation.
For instance, one can start a study with a simple minded local operator for the $\Theta^{+}(1540)$,
which is constructed from the product of a neutron operator and a $K^{+}$ 
operator such as $\Theta = \varepsilon_{abc}(d^{T}_{a}C\gamma_5 u_{b})d_{c}({\bar s}_{e}\gamma_5 u_{e})$.
The two-point correlation function composed of this operator,
in general, couples not only to the $\Theta$ state (single hadron)
but also to the two-hadron states such as 
an interacting $KN$ system~\cite{{Luscher:1986pf},{Fukugita:1994ve}}.
Even worse, when the mass of the $qqqq{\bar q}$ state is
higher than the threshold of the hadronic two-body system, the two-point function
should be dominated by the two-hadron states.
Thus, a specific operator with as
little overlap with the hadronic two-body states 
as possible is desired in order to identify the 
signal of the pentaquark state in lattice QCD.

%
%
\begin{figure*}[t]
\centering
\includegraphics[scale=0.45]{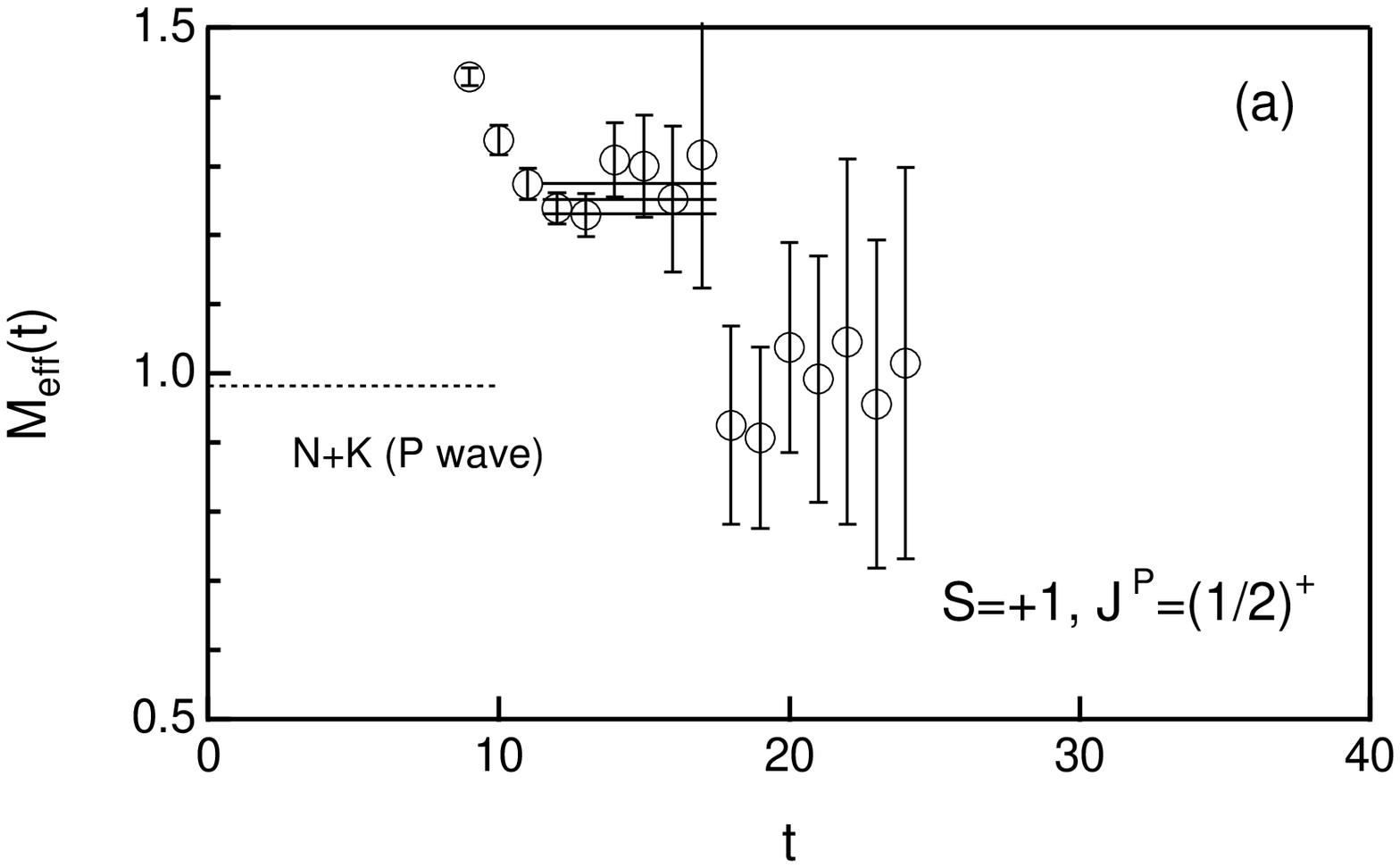}
\includegraphics[scale=0.45]{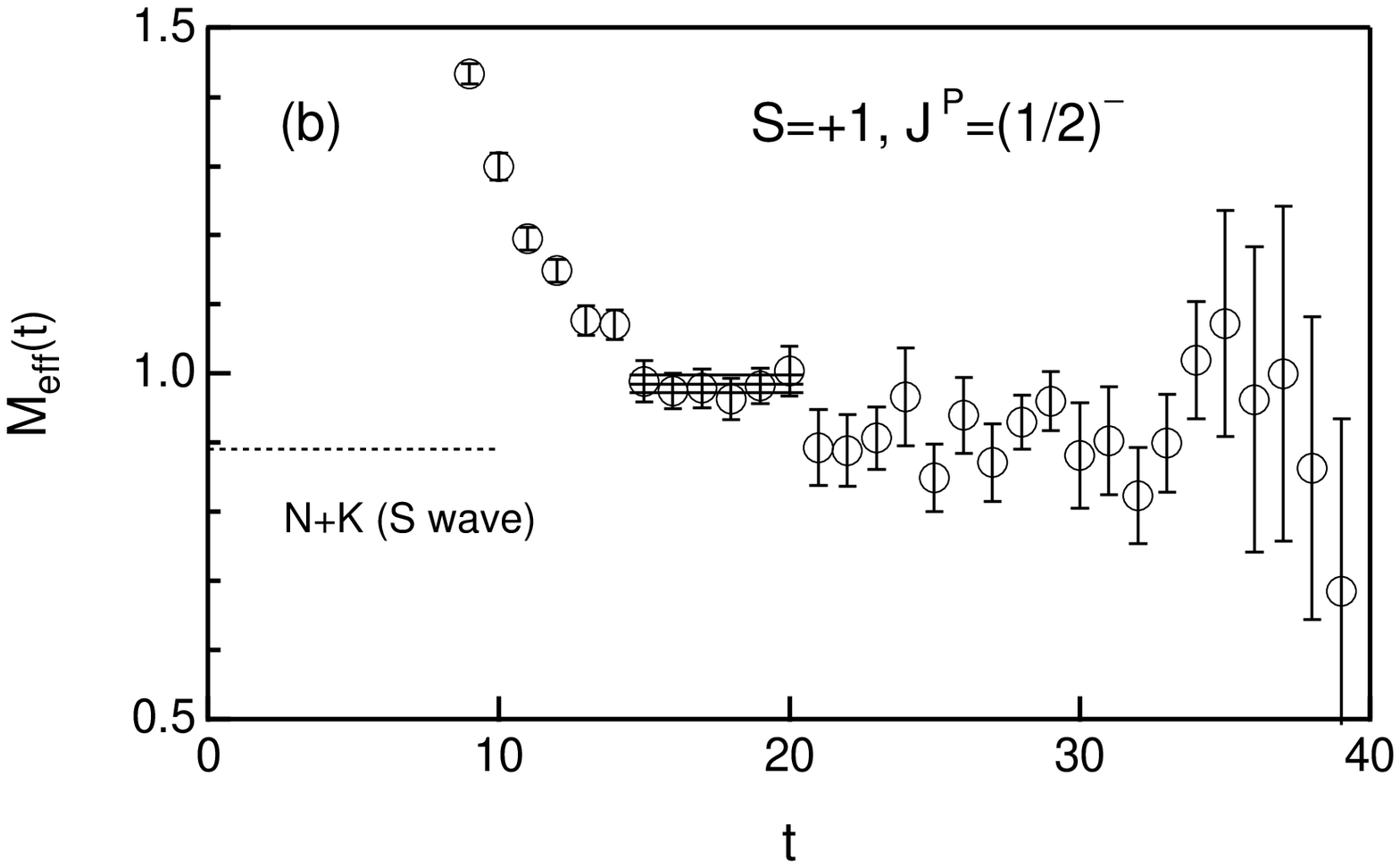}
\caption{Examples of 
the effective mass for (a) the positive parity $\Theta(uudd{\bar s})$ state and (b)
the negative parity $\Theta(uudd{\bar s})$ state at $\kappa=0.1506$ for up and down 
quarks and the strange quark $\kappa_s=0.1515$. The solid lines represent fitted mass and
their statistical errors. The dashed lines correspond to the total energy of 
the non-interacting $KN$ state with the smallest possible nonzero lattice momentum for (a)
and with zero momentum for (b).
}
\label{fig:Theta}
\end{figure*}  

For this purpose, we propose some local interpolating operators
of antidecuplet baryons
based on an exotic description as diquark-diquark-antiquark.
There are basically two choices as ${\bar 3}_{c}\otimes{\bar 3}_{c}$
or ${\bar 3}_{c}\otimes 6_{c}$ to construct a color triplet diquark-diquark 
cluster~\cite{{Carlson:2003pn},{Glozman:2003sy}}. We adopt the former
for a rather simple construction of diquark-diquark-antiquark.
We first introduce the flavor antitriplet (${\bar 3}_{f}$) and 
color antitriplet (${\bar 3}_{c}$) diquark field 
%
%
\begin{eqnarray}
\Phi_{\Gamma}^{i, a}(x)=\frac{1}{2}\varepsilon_{ijk}
\varepsilon_{abc}q_{j,b}^{T}(x)C\Gamma q_{k,c}(x)
\end{eqnarray}
where $C$ is the charge conjugation matrix, $abc$ the color indices, and $ijk$ 
the flavor indices. $\Gamma$ is any of the sixteen Dirac $\gamma$-matrices.
Accounting for both color and flavor antisymmetries, possible $\Gamma$s 
are restricted within $1$, $\gamma_5$ and $\gamma_5\gamma_{\mu}$ 
which satisfy the relation $(C\Gamma)^T=-C\Gamma$. Otherwise, the above defined 
diquark operator is identically zero. 
Hence, three types of flavor ${\bar 3}_{f}$ and color ${\bar 3}_{c}$
diquark; scalar ($\gamma_5$), pseudoscalar ($1$)
and vector ($\gamma_5\gamma_{\mu}$) diquarks
are allowed~\cite{Hadron2003}. 
The color singlet 
state can be constructed by the color antisymmetric parts of diquark-diquark 
$({\bar 3}_{c}\otimes{\bar 3}_{c})_{\rm antisym}={3}_{c}$
with an antiquark (${\bar 3}_{c}$). 
In terms of flavor, ${\bar 3}_{f}\otimes{\bar 3}_{f}\otimes {\bar 3}_{f}={1}_{f}\oplus 8_{f}\oplus 8_{f}\oplus \overline{10}_{f}$. Manifestly, in this description, the $S$=+1 state belongs to the flavor antidecuplet~\cite{Hosaka:2003jv}. Automatically, the $S$=+1 state should have isospin zero. Then, the interpolating operator of the $\Theta(uudd{\bar s})$ is obtained as
%
%
\begin{eqnarray}
\Theta(x)=\varepsilon_{abc} \Phi_{\Gamma}^{s, a}(x)
\Phi_{\Gamma '}^{s, b}(x)C{\bar s}^{T}_{c}(x)
\end{eqnarray}
for $\Gamma \neq \Gamma '$.  The form $C{\bar s}^{T}$
for the strange antiquark field is responsible for 
giving the proper transformation properties of the resulting pentaquark 
operator under parity and Lorentz transformations~\cite{Sugiyama:2003zk}.
Note that because of the color antisymmetry, 
the combination of the same types of diquark is not allowed. 
Consequently, we have three different types of exotic $S$=+1
baryon operators through the combination of two different types of diquarks, which have
different spin-parity~\cite{Hadron2003}:
%
%
\begin{eqnarray}
\label{eq:1stOP}
\Theta^{1}_{+}&=&
\varepsilon_{abc}\varepsilon_{aef}\varepsilon_{bgh}
(u_{e}^{T}Cd_{f})(u_{g}^{T}C\gamma_{5}d_{h})C{\bar s}^{T}_{c} ,\\
\label{eq:2ndOP}
\Theta^{2}_{+, \mu}&=&
\varepsilon_{abc}\varepsilon_{aef}\varepsilon_{bgh}
(u_{e}^{T}C\gamma_{5}d_{f})(u_{g}^{T}C\gamma_{5}\gamma_{\mu}d_{h})C{\bar s}^{T}_{c} ,\\
\label{eq:3rdOP}
\Theta^{3}_{-, \mu}&=&
\varepsilon_{abc}\varepsilon_{aef}\varepsilon_{bgh}
(u_{e}^{T}Cd_{f})(u_{g}^{T}C\gamma_{5}\gamma_{\mu}d_{h})C{\bar s}^{T}_{c} 
\end{eqnarray}
where the subscript $``+(-)"$ refers to positive (negative) parity since
these operators transform as ${\cal P}\Theta_{\pm}({\vec x},t){\cal P}^{\dag}
=\pm\gamma_{4}\Theta_{\pm}(-{\vec x},t)$ (for $\mu = 1,2,3$) under parity.
The first operator of Eq.~(\ref{eq:1stOP}) is proposed for QCD sum rules 
in a recent paper~\cite{Sugiyama:2003zk} independently.

In this description, the operator of exotic $\Xi_{3/2}$
($ssdd {\bar u}$ or $uuss {\bar d}$) states, which are members 
of the antidecuplet, can be treated by interchanging $u$ and $s$
or $d$ and $s$ in the above operators. 
If a strange antiquark is simply replaced by a charm antiquark,
the proposed pentaquark operators can be regarded as
the anti-charmed analog of the isosinglet pentaquark state, $\Theta_c(uudd {\bar c}$).

Recall that any of local type baryon operators can couple to both positive- and 
negative-parity states since the parity assignment of an operator is switched by 
multiplying the left hand side of the operator by $\gamma_5$.  
The desired parity state is obtained by choosing the appropriate projection operator, 
$1\pm \gamma_4$, on the two-point function $G(t)$ and direction of propagation in time. 
Details of the parity projection are described in Ref. \cite{Sasaki:2001nf}. 
We emphasize that the second and third operators, Eqs. (\ref{eq:2ndOP}) and (\ref{eq:3rdOP}),
can couple to both spin-1/2 and spin-3/2 states. 
By using them, it is possible to study the spin-orbit partner of the spin-1/2 $\Theta$ state,
whose presence contradicts the Skyrme picture of the $\Theta$~\cite{Glozman:2003sy}.
However, we will not pursue this direction in this article.
We utilize only the first operator of  Eq.~(\ref{eq:1stOP}), which couples only to a 
spin-1/2 state.  Under the assumption of the highly correlated diquarks,
we simply omit a quark-exchange diagram between diquark pairs
contributing to the full two-point function
in the following numerical simulations.

We generate quenched QCD configurations on a lattice
$L^3\times T=32^3\times48$ with the standard single-plaquette 
Wilson action at $\beta=6/g^{2}=6.2$ ($a^{-1}=2.9$ GeV). 
The spatial lattice size corresponds to $La\approx 2.2 {\rm fm}$,
which may be marginal for treating the ground state of baryons
without large finite volume effect.
Our results are analyzed on 240 configurations.
The light-quark propagators are computed using the Wilson fermions at
four values of the hopping parameter $\kappa=\{0.1520, 0.1506, 0.1497, 0.1489\}$, 
which cover the range $M_{\pi}/M_{\rho}=0.68-$0.90. $\kappa_{s}=0.1515$ and
$\kappa_{c}=0.1360$ are reserved for the strange and charm masses, which are 
determined by approximately reproducing masses of $\phi(1020)$ and $J/\Psi(3097)$. 
We calculate a simple point-point quark propagator with a source location at $t_{\rm src}=6$.
To perform precise parity projection, we construct forward propagating quarks
by taking the appropriate linear combination of propagators with periodic and 
anti-periodic boundary conditions in the time direction. 
This procedure yields a forward in time propagation in the time slice range
$0<t<T-t_{\rm src}$.

%
%
\begin{figure}[t]
\centering
\includegraphics[scale=0.45]{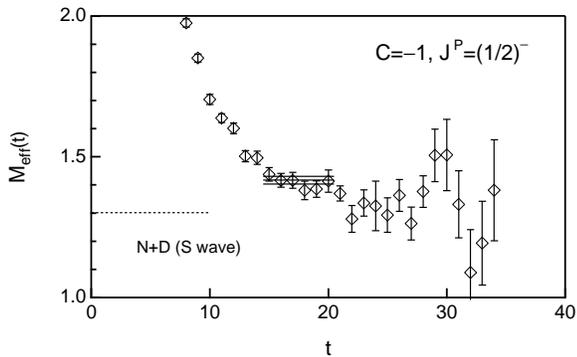}
\caption{
An example of 
the effective mass for the negative parity state
of the charm analog $\Theta_c(uudd{\bar c})$
at $\kappa=0.1506$ for up and down 
quarks and the charm quark $\kappa_c=0.1360$.
}
\label{fig:ThetaC}
\end{figure}  

In this calculation, the strange (charm) quark mass is fixed at $\kappa_s$
($\kappa_c$) and the up and down quark masses are varied from
$M_{\pi}\approx 1.0$ GeV ($\kappa=0.1489$) to $M_{\pi}\approx 0.6$ 
GeV ($\kappa=0.1520$). Then, we perform the extrapolation to the chiral limit
using five different $\kappa$ values. 

We first calculate the effective masses $M_{\rm eff}(t)=\ln\{G(t)/G(t+1)\}$ 
for both parity states of the spin-1/2 $\Theta(uudd{\bar s})$.
For example, Figs.~\ref{fig:Theta}
show effective masses for the positive parity channel and the negative 
parity channel at $\kappa=0.1506$ for up and down quarks with the fixed strange 
quark. 
Statistical uncertainties in both figures are
estimated by a single elimination jack-knife method.

In Fig.~\ref{fig:Theta} (a), the effective mass plot for the positive parity state
shows a very short plateau albeit with large statistical errors.
This plateau terminates at $t\approx 17$ and then the rather 
noisy signals appear after $t=18$ and become reduced around
the $KN$ threshold. We remark that the positive parity $\Theta$ 
state can decay into the $KN$ state in a P-wave where the 
two hadrons should have a nonzero momentum. However, all momenta 
are quantized as ${\vec p}_{n}=2\pi{\vec n}/L$ 
on a system of finite volume. The $KN$ threshold is defined as the total 
energy of the non-interacting $KN$ state with the smallest nonzero 
momentum $|{\vec p}_{\rm min}|=2\pi/L$ in lattice units.
Here, we stress the following two points.
First, there is {\it no clear signal for the $KN$ state} to be observed in 
the effective mass plot. It means that our proposed interpolating operator 
couples weakly to the $KN$ scattering state. Secondly, our observed plateau 
in Fig.~\ref{fig:Theta} (a) is considerably higher than the $KN$ threshold.
While the observed asymptotic state can be identified as a pentaquark (single hadron) 
state, our results seem to give no indication of 
the $\Theta^{+}(1540)$ state in the positive parity channel.

In the negative parity channel,
the gross feature is similar to the case of the positive parity. 
Fig.~\ref{fig:Theta} (b) shows that a clear plateau appears
in the range $15\le t \le 20$. The relatively noisy signals appear 
around the S-wave $KN$ threshold 
after $t=21$ and continue toward the maximum time slice 
$t=T-t_{\rm src}$ for the forwarding propagation.
The errors after $t=21$ are probably underestimated.
The correlators for the heavier mass state in this euclidean 
time region usually have many orders of magnitude deviation 
and the distribution is non-Gaussian. Therefore, the signals
after $t=21$ are inconclusive.

We perform a covariant single exponential fit \footnote{ 
A simple exponential 
might not be an appropriate functional form 
for the decaying state. 
Recall that the pentaquark state can decay into
two-hadron states even in the quenched approximation. Strictly speaking,
we should take the decay width into account in the fitting form. 
However, nobody knows an appropriate analytic form for the two-point function
of an unstable state {\it in finite volume} on the lattice.}
to the two-point function in the  plateau region $15\le t \le 20$,
where the respective $\chi^2$ is indeed most favorable.
The estimated mass is clearly higher than the $KN$ threshold,
which is evaluated as the total energy of the non-interacting 
$KN$ state with zero momentum. 
The excitation energy of the observed asymptotic state from 
the $KN$ threshold is roughly consistent with the experimental value.
Although, without a finite volume analysis, it cannot be excluded that the 
observed plateau stems from only a mixture of the $KN$ scattering states; 
we may regard it as a pentaquark state with a mass close to the    
experimental value of the $\Theta^{+}(1540)$.

As a strange antiquark is simply replaced by a charm antiquark,
we can explore the anti-charmed pentaquark  $\Theta_c$($uudd{\bar c}$)
as well. A similar identification for the $\Theta_c$ state 
can be made in the negative parity channel.
The effective mass plot (Fig.~\ref{fig:ThetaC}) shows that a plateau, 
which terminates at $t\approx21$, is much higher than 
the $DN$ threshold. The relatively noisy signals appear around this 
threshold after $t\approx21$. The observed asymptotic state is identified 
with a pentaquark (single hadron) state similarly.

\begin{figure}[tb]
\centering
\includegraphics[scale=0.45]{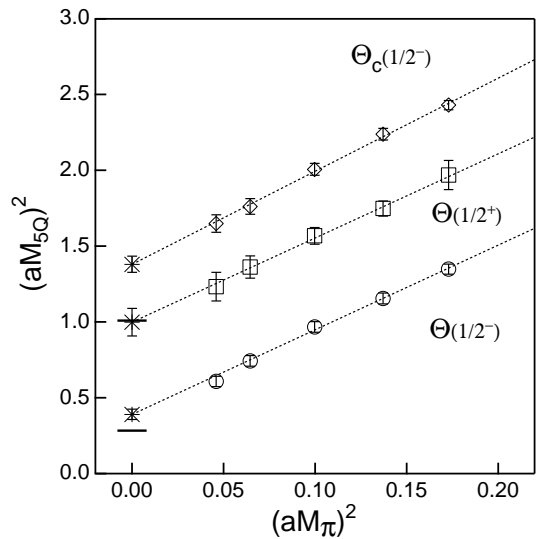}
\caption{
Masses of the spin-1/2 $\Theta(uudd{\bar s})$ states with both positive
parity (open squares) and negative parity (open circles)
as functions of pion mass squared in lattice units.
The charm analog  
$\Theta_c(uudd{\bar c})$ state
(open diamonds) is also plotted.
Horizontal short bar represents the $KN$($DN$) threshold estimated 
by $M_{N}+M_{K}$($M_{N}+M_{D}$) in the chiral limit.
}
\label{fig:Theta5Q}
\end{figure}

In Fig.~\ref{fig:Theta5Q} we show the mass spectrum of the 
$\Theta(uudd{\bar s})$ states with the positive parity (open squares) 
and the negative parity (open circles) as functions of the pion mass squared. 
Mass estimates are obtained from covariant single exponential fits
in the appropriate fitting range.
All fits have a confidence level larger than 0.3 and $\chi^{2}/N_{DF}<1.2$.
It is evident that the lowest state of the isosinglet $S$=+1 baryons 
has {\it the negative parity}. 
We evaluate the mass of 
the $\Theta(uudd{\bar s})$ with both parities in the chiral limit.
A simple linear fit for all five values in 
Fig.~\ref{fig:Theta5Q} yields 
$M_{\Theta(1/2^-)}$=0.62 (3) and $M_{\Theta(1/2^+)}$=1.00 (5) in lattice units.
If we use the scale set by $r_0$ from Ref.~\cite{Necco:2001xg}, 
we obtain $M_{\Theta(1/2^-)}=1.84 (8)$ GeV and 
$M_{\Theta(1/2^+)}=2.94 (13)$ GeV.  It is worth quoting other related 
hadron masses. The chiral extrapolated values for the kaon, the nucleon and the 
$N^*$ state are $M_{K}=0.53 (1)$ GeV, $M_{N}=1.06 (2)$ GeV and 
$M_{N^*}=1.76 (5)$ GeV in this calculation.

Our obtained $\Theta(1/2^-)$ mass is slightly overestimated
in comparison to the experimental value of the $\Theta^{+}(1540)$, 
but comparable to our observed $N^*$ mass, which is also overestimated.
Needless to say, the evaluated values should not be taken too seriously 
since they do not include any systematic errors. 
Such a precise quantitative prediction of hadron masses is not the 
purpose of the present paper. 
Rather, we emphasize that, our results strongly indicate
the $J^P$ assignment of the $\Theta^{+}(1540)$  
is most likely $(1/2)^-$.
This conclusion is consistent with 
that of a recent lattice study~\cite{Csikor:2003ng}
(if one corrects the parity assignment of their 
operator~\cite{Fodor}) and that of QCD sum rules 
approach~\cite{Sugiyama:2003zk}.

Results for the lowest-lying 
spin-1/2 $\Theta_c$ state, which has the negative parity, are also included
in Fig.~\ref{fig:Theta5Q}. The $\Theta_c$ state lies much higher than the 
$DN$ threshold in contradiction with
several model predictions~\cite{{Jaffe:2003sg},{Stancu:1998sm},{Karliner:2003si}}.
The chiral extrapolated value of the $\Theta_c$ mass is 
3.45(7) GeV, which is about
500 MeV above the $DN$ threshold ($M_{D}$=1.89(1) GeV) 
in our calculation. This indicates 
that the anti-charmed pentaquark $\Theta_c$ 
is not to be expected as a bound state.

%
%

We have calculated the mass spectrum of the $S$=+1 exotic baryon,
$\Theta(uudd{\bar s})$, and the charm analog $\Theta_c(uudd{\bar c})$ 
in quenched lattice QCD.
To circumvent the contamination from hadronic two-body states, 
we formulated the antidecuplet baryon interpolating operators 
using an exotic description like diquark-diquark-antiquark. 
Our lattice simulations seem to give no indication of a pentaquark 
in the positive parity channel to be identified with the $\Theta^+(1540)$.
In contrast the simulations in the negative parity channel 
can easily accommodate a pentaquark with a mass
close to the experimental value. 
Although more detailed lattice study
would be desirable to clarify the significance of this observation, 
the present lattice study favors
spin-parity $(1/2)^-$ for the $\Theta^+(1540)$.
We have also found that the lowest spin-1/2 $\Theta_c$ state,
which has the negative parity, lies much higher than the $DN$ threshold,
in contrast to several model predictions~\cite{{Jaffe:2003sg},{Stancu:1998sm},
{Karliner:2003si}}.

To establish the parity of the $\Theta^+(1540)$, 
more extensive lattice study is required. 
Especially, a finite volume analysis is necessary to
disentangle the pentaquark signal from a mixture of 
the $KN$ scattering states.
It is also important to explore the chiral limit.
This calculation was performed using relatively heavy 
quark mass so that one may worry about a level switching between both
parity states toward the chiral limit as observed in the case 
of excited baryons \cite{{Sasaki:2003xc},{Dong:2003zf}}.
We remark that a study for the non-diagonal correlation between 
our pentaquark operators and a standard two-hadron operator should 
shed light on the structure of the very narrow resonance $\Theta^{+}(1540)$. 
The possible spin-orbit partner of the $\Theta$ state is also accessible 
by using two of our proposed operators.
We plan to further develop the present calculation to involve
more systematic analysis and more detailed discussion.

\begin{acknowledgments}
It is a pleasure to acknowledge A. Hosaka, T. Nakano and L. Glozman
for useful comments. I would also like to thank T. Doi, M. Oka and T. Hatsuda
for fruitful discussions on the subject to determine the parity of the $\Theta$ state,
and S. Ohta for helpful suggestions and his careful reading of the manuscript.
This work is supported by the Supercomputer Project No.102 
 (FY2003) of the High Energy Accelerator Research Organization (KEK)
 and JSPS Grant-in-Aid for Encouragement of Young Scientists (No. 15740137).
\end{acknowledgments}
 

\end{document}